\documentclass[conference]{IEEEtran}
\IEEEoverridecommandlockouts

\usepackage{url}
\usepackage[whole]{bxcjkjatype}
\usepackage{booktabs}

\usepackage{tikz}

\usepackage{stfloats}

\makeatletter
\newcommand\notsotiny{\@setfontsize\notsotiny\@vipt\@viipt}%
\makeatother

\newcommand{\blackcircled}[2][1ex]{%
  \tikz[baseline=(char.base)]{
    \node[shape=circle,fill=black,inner sep=0pt,minimum size=#1,text=white] (char) {\ensuremath{#2}};
  }%
}

\usepackage{caption}
\usepackage{cite}
\usepackage{amsmath,amssymb,amsfonts}
\usepackage{algorithmic}
\usepackage{graphicx}
\usepackage{textcomp}
\usepackage{xcolor}
\def\BibTeX{{\rm B\kern-.05em{\sc i\kern-.025em b}\kern-.08em
    T\kern-.1667em\lower.7ex\hbox{E}\kern-.125emX}}

\usepackage{listings}
\usepackage{xcolor}

\lstdefinelanguage{json}{
    basicstyle=\ttfamily\small,
    numbers=left,
    numberstyle=\scriptsize,
    stepnumber=1,
    numbersep=8pt,
    showstringspaces=false,
    breaklines=true,
    frame=single,
    backgroundcolor=\color{white},
    stringstyle=\color{red!60!black},
    keywordstyle=\color{blue},
    morestring=[b]",
    morestring=[d]',
    literate=
     *{0}{{{\color{orange}0}}}{1}
      {1}{{{\color{orange}1}}}{1}
      {2}{{{\color{orange}2}}}{1}
      {3}{{{\color{orange}3}}}{1}
      {4}{{{\color{orange}4}}}{1}
      {5}{{{\color{orange}5}}}{1}
      {6}{{{\color{orange}6}}}{1}
      {7}{{{\color{orange}7}}}{1}
      {8}{{{\color{orange}8}}}{1}
      {9}{{{\color{orange}9}}}{1}
      {:}{{{\color{blue}:}}}{1}
      {,}{{{\color{blue},}}}{1}
      {\{}{{{\color{blue}\{}}}{1}
      {\}}{{{\color{blue}\}}}}{1}
      {[}{{{\color{blue}[}}}{1}
      {]}{{{\color{blue}]}}}{1},
}

\usepackage{eso-pic}

\AddToShipoutPictureBG*{
  \AtPageCenter{
    \raisebox{12.8cm}{
      \makebox[0pt]{
        \fbox{\parbox{0.95\textwidth}{\small This paper has been accepted for publication at IEEE Globecom 2025. If you wish to cite this work, please use the following reference: Hiroki Nakano, Takashi Koide, and Daiki Chiba, ``PhishParrot: LLM-Driven Adaptive Crawling to Unveil Cloaked Phishing Sites,'' Proceedings of IEEE Global Communications Conference (GLOBECOM), 2025.}}
      }
    }
  }
}

\begin{document}

\title{PhishParrot: LLM-Driven Adaptive Crawling \\to Unveil Cloaked Phishing Sites}

\author{
    \IEEEauthorblockN{Hiroki Nakano\IEEEauthorrefmark{1}, Takashi Koide\IEEEauthorrefmark{1}, Daiki Chiba\IEEEauthorrefmark{1}}
    \IEEEauthorblockA{\IEEEauthorrefmark{1}NTT Security Holdings Corporation \& NTT, Inc., Japan\\
    Email: hi.nakano.sec@gmail.com}
}

\maketitle

\begin{abstract}
Phishing attacks continue to evolve, with cloaking techniques posing a significant challenge to detection efforts.
Cloaking allows attackers to display phishing sites only to specific users while presenting legitimate pages to security crawlers, rendering traditional detection systems ineffective.
This research proposes PhishParrot, a novel crawling environment optimization system designed to counter cloaking techniques.
PhishParrot leverages the contextual analysis capabilities of Large Language Models (LLMs) to identify potential patterns in crawling information, enabling the construction of optimal user profiles capable of bypassing cloaking mechanisms.
The system accumulates information on phishing sites collected from diverse environments.
It then adapts browser settings and network configurations to match the attacker's target user conditions based on information extracted from similar cases.
A 21-day evaluation showed that PhishParrot improved detection accuracy by up to 33.8\% over standard analysis systems, yielding 91 distinct crawling environments for diverse conditions targeted by attackers.
The findings confirm that the combination of similar-case extraction and LLM-based context analysis is an effective approach for detecting cloaked phishing attacks.

\end{abstract}

\begin{IEEEkeywords}
Phishing, Cloaking, Large Language Model.
\end{IEEEkeywords}

\section{Introduction}

 \begin{figure*}[!t]
     \centering
         \includegraphics[scale=0.5]{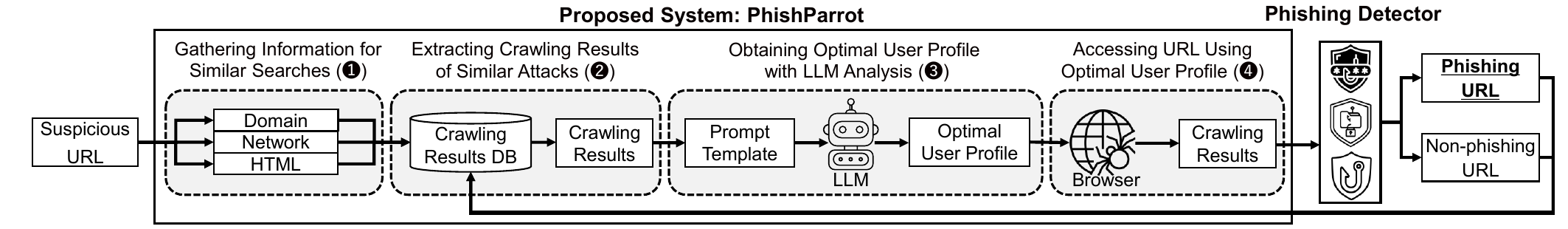}
         \caption{PhishParrot is composed of 4 steps to overcome sophisticated cloaking techniques and improve phishing detection accuracy. The system takes suspicious URLs as input and outputs actual phishing content hidden by cloaking mechanisms. The system performs a preliminary access to gather information for similar attack searches (\blackcircled{1}). It extracts crawling results of similar attacks from a database using cloaking pattern similarity-based retrieval (\blackcircled{2}). It embeds this information into prompts and requests LLM analysis to determine the optimal user profile matching attacker's targeting criteria (\blackcircled{3}). It configures the browser with this profile and accesses the URL to retrieve phishing content hidden from security crawlers (\blackcircled{4}). This content is sent to phishing detectors, enabling accurate classification impossible with conventional crawling approaches. Without PhishParrot's adaptive approach, even sophisticated phishing detectors would fail as cloaking mechanisms prevent access to malicious content.}
         \label{fig:phishParrot}
 \end{figure*}

Phishing attacks have become increasingly sophisticated, posing a serious threat to internet users worldwide.
Attackers employ ``cloaking'' techniques to evade detection mechanisms.
They display phishing content only to users who meet specific criteria, while presenting legitimate pages or error messages to security crawlers and other unwanted visitors~\cite{DBLP:conf/sp/ZhangOCSJWSKBWS21,DBLP:conf/ccs/ZhangSKBBCO0BSA22}.

Conventional phishing detection systems analyze features such as HTML content and screenshots using machine learning models or rule-based algorithms, with some incorporating network data like IP addresses and HTTP transactions~\cite{DBLP:conf/ccs/AbdelnabiKF20,DBLP:conf/uss/LinLDNCLSZD21,DBLP:conf/uss/Liu0YNDD22,DBLP:journals/access/KoideNC24}.
However, these systems fundamentally depend on successfully retrieving phishing content. When attackers identify security crawlers through cloaking mechanisms, they serve non-phishing content instead, which makes accurate analysis impossible.
Research analyzing security crawler profiling has demonstrated that even when simulating ``typical user-like'' requests, without appropriate environmental adjustments, crawlers cannot satisfy all conditions targeted by attackers~\cite{DBLP:conf/uss/AcharyaV21}.
Consequently, content acquisition methods dependent on fixed crawling environments have limited ability to reach phishing content.
To address these limitations, a new approach capable of flexibly and accurately capturing the characteristics of phishing attacks is essential.

This research proposes PhishParrot, a novel crawling environment optimization system designed to improve detection rates for cloaked phishing sites.
PhishParrot adaptively adjusts browser profiles (such as HTTP headers) and network environments (e.g., country information, residential or mobile networks) based on crawling results information to satisfy attacker-defined conditions and access phishing content (Fig.~\ref{fig:phishParrot}).
The system accumulates phishing site information (Domain Information, Network Information, HTML Information) collected from diverse environments and incorporates a mechanism to extract similar cases.
Based on this information, a Large Language Model (LLM) suggests optimal user profiles (browser settings and access environments), enabling the construction of environments closely resembling those targeted by attackers, thereby bypassing cloaking to retrieve phishing content.
The advanced language comprehension capabilities of LLMs enable the flexible and accurate identification of trending characteristics in the latest phishing attacks.
This potentially enhances effectiveness against cloaking-based detection avoidance, a challenge for conventional fixed-environment crawling approaches.

In this study, we evaluated PhishParrot's performance using three types of LLMs on 32,487 suspicious URLs collected over a 21-day period.
Results showed that the best-performing model achieved 88.0\% accuracy with an average processing time of 16.69 seconds per URL.
When combining PhishParrot with three different Phishing Detectors, we confirmed up to a 33.8\% improvement in phishing site detection rates compared to standard analysis systems.
PhishParrot also demonstrated the ability to suggest 91 optimal user profiles based on attack trends and phishing site similarities, enabling flexible environment construction tailored to attackers' targets.
These results indicate that rule-based approaches struggle to comprehensively cover the diverse user access environments targeted by attackers.
In contrast, the sophisticated context analysis capabilities of LLMs enable flexible and effective crawling environment construction unachievable by other approaches.
The findings of this research demonstrate that incorporating LLMs into systems is an effective strategy for addressing the challenge of detection avoidance through cloaking.

\section{Related Work}
Numerous studies have been conducted on phishing site detection, with many systems proposed to identify phishing sites based on HTML content and screenshots obtained when accessing URLs~\cite{DBLP:conf/ccs/AbdelnabiKF20,DBLP:conf/uss/Liu0ZLD23,DBLP:journals/access/KoideNC24,DBLP:journals/compsec/ChibaNK25}.
Liu et al. proposed ``PhishIntention,'' which visually extracts both brand intention and credential-harvesting intention from webpage appearance and dynamic behavior, verifying credential-harvesting intention through interaction with the webpage~\cite{DBLP:conf/uss/Liu0YNDD22}.
Lin et al. introduced ``Phishpedia,'' a system that accurately recognizes brand logos from webpage screenshots and matches logo variations of the same brand, enabling high-precision identification and visual explanation of phishing sites~\cite{DBLP:conf/uss/LinLDNCLSZD21}.
While these studies demonstrate high detection capabilities in situations where phishing content is accessible, their detection accuracy inevitably decreases when accessing continuously from the same or known analysis environments, as cloaking techniques prevent observation of the actual phishing content.

Cloaking techniques refer to technologies that display different content based on the characteristics of the access source.
When a phishing URL is accessed, attackers analyze request characteristics in detail (User-agent, IP address, HTTP headers, etc.).
Based on this analysis, they distinguish between ``Victim Access'' (access from targeted typical users) and ``Non-victim Access'' (access from security researchers or detection systems).
According to Zhang et al.~\cite{DBLP:conf/ccs/ZhangSKBBCO0BSA22}, 96.52\% (2,831 cases) of the 2,933 phishing kits analyzed contained cloaking techniques.
IP address and User-agent based cloaking were found to be the most commonly used, with 1,983 cases (approximately 82\%) implementing User-Agent verification and 1,660 cases (approximately 69\%) implementing IP address verification among the 2,421 phishing kits analyzed.

To address these cloaking techniques, researchers have proposed analyses of cloaking technologies embedded in phishing kits and techniques to bypass specific types of cloaking~\cite{DBLP:conf/sp/InvernizziTKCPB16,DBLP:conf/ccs/ZhangSKBBCO0BSA22,DBLP:conf/uss/Teoh0LHD24}.
Zhang et al. proposed ``CrawlPhish,'' a framework for automatically detecting and classifying client-side cloaking used in phishing sites~\cite{DBLP:conf/sp/ZhangOCSJWSKBWS21}.
Their analysis of 112,005 phishing sites collected over 14 months from 2018 to 2019 revealed that the use of cloaking techniques increased from 23.32\% to 33.70\%. 
Acharya et al. proposed the ``PhishPrint'' framework, which pre-profiles the characteristics of security crawlers to identify weaknesses and uses them to evade phishing detection through cloaking attacks~\cite{DBLP:conf/uss/AcharyaV21}.
Through evaluation of 23 security crawlers, they revealed weaknesses against cloaking, such as a lack of diversity in browser fingerprints and IP address limitations.
While existing research implements simple countermeasures against cloaking, our proposed system differs significantly in that it adapts to the latest phishing attack trends and constructs access environments that more effectively bypass cloaking.

\section{Proposed System: PhishParrot}

\subsection{Design Goals and Objectives}
To bypass attackers' cloaking techniques and retrieve phishing content that typical users encounter, we propose ``PhishParrot'' (Fig.~\ref{fig:phishParrot}).
Our system focuses primarily on attributes such as HTTP headers and IP addresses, which constitute the majority of current cloaking techniques, to accurately reproduce the target user's access environment.
The system aims to achieve the following two goals:

\noindent\textbf{Goal 1: Identify similar cloaking patterns.}
Phishing attacks have been reported to share similarities based on toolkits or attacker groups~\cite{DBLP:conf/ccs/ZhangSKBBCO0BSA22}.
Cloaking techniques follow similar trends; therefore, we extract similar examples from accumulated crawling results.

\noindent\textbf{Goal 2: Bypass cloaking to access phishing content.}
Bypassing cloaking techniques requires mimicking ``typical user-like'' target access.
Based on past successful examples, LLM analyzes appropriate crawling environments and retrieves content from target URLs.

\subsection{Gathering Information for Similar Searches (\blackcircled{1})}
\begin{table}[!t]
\notsotiny
\tabcolsep=0.6mm
\centering
\caption{Defined Structure of Crawling Results}
\begin{tabular}{lll} \toprule
Category & Type & Content \\ \midrule
Domain & Registration & \texttt{\{"domain\_name": "EXAMPLE.COM", }\\
Information & & \quad\texttt{"registrar": "...", ...\}}\\
& DNS & \texttt{\{"Status": 0, "Answer": [\{"name": "example.com",} \\
& & \quad\texttt{"TTL: 123, "data": 192.0.2.0", ...\}, \{...\}]\}} \\
 & TLS/SSL & \texttt{\{"issuer": \{"O": "ExampleCert", ...\}, } \\
  &  & \texttt{\quad"subject": \{"CN": "example.com"\}, ...\}} \\
\midrule
Network & Requests & \texttt{\{"requests": [\{"id": "1", "method": "GET", } \\
Information & & \quad\texttt{"url": "https://example.com", "headers":\{...\},} \\
 & & \qquad\texttt{...\}, \{"id": "2", "method": "GET",...\},...]\}} \\ 
  & Responses & \texttt{\{"responses": [\{"id": "1", "status": "200", } \\ 
  &  & \quad\texttt{"url": "https://example.com", "headers": \{...\}} \\ 
 &  & \quad\texttt{,...\{"id": "2", "status": "200",..\},...]\}} \\ 
\midrule
HTML & Visible Text & \texttt{403 Forbidden nginx/1.26.2 You do not} \\
Information & & \texttt{have permission to access this resource.} \\
 & Tag Structure & \texttt{<html><head><title></title><style></style>} \\
 & & \texttt{</head><body><h1></h1><p></p></body></html>} \\
\midrule
Crawling & IP Geolocation & \texttt{\{"country": "US", } \\
Environment & & \quad\texttt{"city": "Sample City", "region":"Sample"\}} \\
Information & ASN & \texttt{\{"asn": "AS1234", "name": "Example Business", } \\
 & & \quad\texttt{"domain": "examplebusiness.example", ...\}} \\
 & Language & \texttt{en-US} \\
  \bottomrule
\end{tabular}
\label{tab:structure_define}
\end{table}

In this step, we define the data structure of crawling results used for similar case searches and LLM prompt embedding.
As shown in Table~\ref{tab:structure_define}, we establish data structures in four categories: Domain Information, Network Information, HTML Information, and Crawling Environment Information, and collect this information.

\noindent\textbf{Domain Information.}
In some cases, domain names can be analyzed to reveal characteristic strings or subdomain structures reused from past phishing attacks~\cite{DBLP:conf/acsac/KoideFN023}.
Additionally, registration timing and certificate reuse can help identify attacks from the same group.
If similar attack cases can be identified, cloaking techniques may also exhibit similar patterns, enabling the construction of optimal user profiles based on past successful cases.
Therefore, we extract domain names from target URLs and obtain registration information, DNS resolution, and TLS/SSL certificates.

\noindent\textbf{Network Information.}
When cloaking techniques exist in a target URL, characteristic responses similar to past cases may occur.
For example, security crawler access has been reported to be redirected to legitimate sites of impersonated companies or prominent sites (e.g., Google homepage)~\cite{DBLP:conf/asiaccs/0004HK24}.
To capture this trend, we perform a simple access to the target URL using Playwright and collect requests and responses.

\noindent\textbf{HTML Information.}
Phishing sites often mimic prominent companies and copy legitimate sites, resulting in similar HTML structures and appearances~\cite{DBLP:conf/uss/LinLDNCLSZD21,DBLP:conf/ccs/AbdelnabiKF20}.
By analyzing the similarity between HTML and past phishing attack crawling results, it becomes possible to identify identical cloaking patterns.
From the DOM tree obtained during the simple access mentioned above, we extract visible text and HTML tag structures.

\noindent\textbf{Crawling Environment Information.}
Attackers have been reported to determine routing destinations based on network information from the access source~\cite{DBLP:conf/sp/InvernizziTKCPB16,DBLP:conf/ccs/ZhangSKBBCO0BSA22}.
For example, cloaking techniques may be applied to IP addresses that appear to be from analysis environments or accesses from countries other than the target, preventing the retrieval of phishing content.
Conversely, access from general residential or mobile networks in specific countries may match the attacker's target and reach the phishing site.
Therefore, we store network location (IP Geolocation), network provider (ASN), and browser language settings (Language) for each crawling result.

\subsection{Extracting Crawling Results of Similar Attacks (\blackcircled{2})}

In this step, we extract crawling results of similar attacks using state-of-the-art vector representation techniques and search methods.

\noindent\textbf{Crawling Results DB Integration.}
We store crawling results (with the defined data structure) in a vector database, labeling them as successful crawling or failed crawling.
Vector databases enable comparisons based on semantic similarity, allowing effective retrieval of past crawling results from similar attacks.
For string vectorization, we utilize OpenAI's text-embedding-3-small model, which transforms text into a 1,536-dimensional vector space.
We selected this model because it offers an optimal balance between analysis speed and performance compared to other prominent models (text-embedding-3-large, text-embedding-ada-002).
While the system requires a certain amount of labeled initial data to operate, as shown in Fig.~\ref{fig:phishParrot}, it automatically expands successful crawling and failed crawling examples by combining continuous system operation with detection results from Phishing Detectors.

\noindent\textbf{Similarity-based Crawling Results Extraction.}
In PhishParrot, we convert information from the three defined categories (Domain, Network, and HTML Information) into string representations and extract similar crawling results based on semantic similarity.
For example, when searching Domain Information, we compare similarity using 1,536-dimensional vectors of registration, DNS, and TLS/SSL strings.
The procedure for extracting crawling results is as follows:
\begin{enumerate}
\item Create embedding vectors for each Domain Information, Network Information, and HTML Information of the target URL

\item Retrieve results with cosine similarity above a certain threshold from successful and failed examples in the database

\item Generate embedding vectors from the entire text of the results and extract representative results using the Maximum Marginal Relevance (MMR) algorithm~\cite{DBLP:conf/sigir/CarbonellG98}
\end{enumerate}

Cosine similarity quantifies the structural and semantic similarity of target URL information in multidimensional space, enabling high-precision identification of similar patterns in attack methods and strategies.
Cosine similarity ranges from 0 (unrelated) to 1 (identical), and we set a threshold of 0.65 based on preliminary experiments that successfully identified similar attacks.
The MMR algorithm balances relevance and diversity by selecting documents that maximize marginal relevance, creating a representative result set that covers diverse environmental conditions useful for detecting similar cloaking patterns.
While λ in MMR ranges from 0 (emphasizing diversity) to 1 (emphasizing relevance), we adopt 0.7 to prioritize relevance while still considering diversity.

\noindent\textbf{Prompt-Optimized Crawling Results Filtering.}
We filter redundant information from the extracted results for LLM prompts. 
Specifically, we extract only registration from Domain Information, main communications from Network Information, and Crawling Environment Information from each result.
While other information is important for vector comparison, it is unnecessary for LLM's user profile decisions.
Although the number of results for prompt input is arbitrary, we adopt a total of 10 cases: 5 successful examples and 5 failed examples.
From preliminary experiments, we confirmed this number allows the LLM to analyze success (e.g., geolocations) and failure (e.g., desktop restrictions) factors.

\subsection{Obtaining Optimal User Profile with LLM Analysis (\blackcircled{3})}

In this step, we derive the optimal user profile using LLM analysis of the extracted crawling results.

\noindent\textbf{Prompt Definition.}
Table~\ref{tab:prompt_template} shows the prompt for LLM interaction.
Prompts are information provided by users to help LLMs generate appropriate responses.
We define a cybersecurity expert persona specializing in phishing detection and anti-cloaking techniques (Persona).
We provide instructions focusing on attacker strategies for cloaking analysis (Instruction).
We specify that analysis results should be output in JSON format with five types of information (Output Format).
Finally, we embed the target information in the prompt (Analysis Request).

\noindent\textbf{LLM Analysis Request.}
We embed the target URL information as \texttt{url}, successful examples from extracted crawling results as \texttt{successful\_examples}, and failed examples as \texttt{failed\_examples} in the prompt.
This approach, known as Retrieval-Augmented Generation (RAG), enhances LLM reasoning with external knowledge, enabling responses based on current and specific information~\cite{DBLP:conf/nips/LewisPPPKGKLYR020}.
Through RAG, we provide the LLM with phishing site cloaking patterns and past crawling results, facilitating more adaptive crawling environment construction.
The analysis results include three types of information for crawling (\texttt{http\_header}, \texttt{ip\_location}, \texttt{network\_provider}) and two types of rationale information (\texttt{target\_victim}, \texttt{reason}).
In this step, these five types of information constitute the output.

\subsection{Accessing URL Using Crawling Environment (\blackcircled{4})}
We aim to bypass cloaking and reach phishing content using the optimal user profile output by the LLM.
Our system employs Playwright for network log acquisition and operational stability.
We configure headers with those specified in \texttt{http\_header}.
For networking, we utilize multiple pre-prepared environments, setting country information and network types (e.g., typical residential, mobile networks) to match those specified in \texttt{ip\_location} and \texttt{network\_provider}.
We access the target URL in this environment and obtain the final page content, screenshots, and network logs after JavaScript interpretation, which constitute the system's final output.

\begin{table}[!t]
\notsotiny
\tabcolsep=1.0mm
  \centering
    \caption{Prompt for Obtaining Optimal User Profile}
    \begin{tabular}{ll} \toprule
      Tactics & Prompt \\ \midrule
      Persona & You are a cybersecurity analyst specializing in phishing detection and anti-cloaking \\ 
      (System) & techniques. Your role is to analyze website behavior patterns and provide \\
       & recommendations for optimal crawling environments to detect malicious sites \\ 
       & that use cloaking techniques. \\
      \midrule
       Instruction & When analyzing cloaking behavior, you should: \\ 
       (System) & 1. Consider geographical targeting patterns and past attack trends.\\
       & 2. Evaluate HTTP header configurations, including User-Agent strings. \\
       & 3. Assess network provider characteristics, particularly those commonly used by \\
       & \quad attackers. \\
       &  4. Compare provided cloaking and non-cloaking examples to identify behavioral\\
       & \quad differences. \\
       & Additionally, prioritize settings that reflect environments attackers typically target \\
       & based on historical evidence of their tactics, techniques, and procedures (TTPs). \\
       & Incorporate insights into network providers, IP ranges, and User-Agent configurations \\
       & most likely to bypass cloaking mechanisms. \\
       \midrule
       Output & Your responses should be provided in JSON format with specific keys: \\ 
        Format & - http\_header: HTTP headers for optimal browsing environment. \\
        (System) & - ip\_location: IP location of optimal browsing environment. \\ 
        & - network\_provider: Optimal browsing environment network provider. \\
        & - target\_victim: Nature of users targeted by attackers. \\
        & - reason: Logical explanation for the recommendations. \\
        & DO NOT include markdown, code blocks, or any text outside the JSON. \\
        \midrule
        Analysis & Analyze the following website information for potential cloaking behavior and \\
        Request & recommend an optimal crawling environment. \\
        (User) & URL: \texttt{\{url\}} \\
         & Reference Examples: \\
         & 1. Successful Crawling Example (examples of successful accesses): \\
         & \texttt{\{successful\_examples\}} \\
         & 2. Failed Crawling Example (examples of sites using cloaking techniques): \\ 
         & \texttt{\{failed\_examples\}} \\
         & Based on this information and the reference examples, provide recommendations for: \\
         & 1. The necessary browsing environment to avoid being cloaked \\
         & 2. Optimal header information (e.g. User-Agent) for HTTP requests \\
         & 3. Source IP address characteristics (location and network provider) \\
         & Please format your response as JSON with the specified keys: \\
         & target\_victim, http\_header, ip\_location, network\_provider, and reason. \\
       \bottomrule
    \end{tabular}
  \label{tab:prompt_template}
\end{table}

\section{Evaluation}
\subsection{Experimental Setup}
We evaluate the detection accuracy and execution time of PhishParrot for identifying phishing sites.

\noindent\textbf{Baseline Systems.}
Existing cloaking analysis frameworks like CrawlPhish~\cite{DBLP:conf/sp/ZhangOCSJWSKBWS21} and PhishPrint~\cite{DBLP:conf/uss/AcharyaV21} are designed for research purposes to understand cloaking mechanisms, rather than for operational detection.
Therefore, we evaluate PhishParrot against practical baseline systems.
To enable this comparison, we prepared two baseline systems: the Standard Analysis System and the Typical User System.
The Standard Analysis System simulates a typical analysis environment by accessing URLs using Google Chrome in headless mode on Linux, from a cloud environment in the United States.
In contrast, the Typical User System randomly selects one combination of \texttt{http\_header}, \texttt{ip\_location}, and \texttt{network\_provider} to emulate a typical user environment.
Specifically, we defined \texttt{http\_headers} corresponding to 17 different access environments (with Safari restricted to macOS and iOS) by combining five OS types (Windows, macOS, Linux, Android, iOS) and four browsers (Microsoft Edge, Google Chrome, Safari, Mozilla Firefox).
Additionally, to capture high internet usage and regional diversity, we set up 10 \texttt{ip\_locations} (United States, India, United Kingdom, Germany, Japan, Brazil, Saudi Arabia, Canada, Australia, South Korea).
For each \texttt{ip\_location}, we specified three \texttt{network\_providers} (Datacenter, Residential, Mobile), resulting in \(17 \times 10 \times 3 = 510\) total combinations.
We used these two systems as baselines for our comparative evaluation with PhishParrot.

\noindent\textbf{Phishing Detectors.}
To evaluate our system against various detection techniques, we reimplemented three state-of-the-art phishing detectors: ChatPhishDetector~\cite{DBLP:journals/access/KoideNC24}, VisualPhishnet~\cite{DBLP:conf/ccs/AbdelnabiKF20}, and StackModel~\cite{DBLP:journals/fgcs/LiYCYL19}.
We chose these systems because they employ distinct approaches: ChatPhishDetector leverages LLM-based contextual analysis of HTML and screenshots; VisualPhishnet performs visual similarity analysis on screenshots using a triplet convolutional neural network; and StackModel combines URL and HTML features with stacked machine learning models. These diverse perspectives provide a comprehensive evaluation of our system's effectiveness.

\begin{table}[!t]
\notsotiny
\tabcolsep=2.0mm
\centering
\caption{Performance Comparison of PhishParrot by LLM}
\begin{tabular}{lrrrrrrr} \toprule
 & \multicolumn{5}{c}{Avg. of 3 Phishing Detectors} & & \\ 
\cmidrule(lr){2-6}
LLM & Acc & TPR & TNR & Pre & F1 & Execution Time & Cost \\ \midrule
o3-mini & 87.7\% & 89.6\% & 85.0\% & 87.9\% & 0.886 & 26.54 seconds & \$0.016 \\
GPT-4o & 85.9\% & 87.9\% & 83.3\% & 85.9\% & 0.868 & 19.13 seconds & \$0.027 \\
GPT-4o mini & 88.0\% & 90.5\% & 85.3\% & 87.4\% & 0.889 & 16.69 seconds & \$0.002 \\
   \bottomrule
 \end{tabular}
 \label{tab:performance_llm}
 \end{table}

\noindent\textbf{Input Suspicious URLs.}
We collected suspicious URLs from three feeds: CrowdCanary~\cite{DBLP:conf/IEEEares/Nakano0KFYHYM23}, which collects phishing URL reports from X using hashtags like ``\#phishing''; Urlscan.io URLs tagged as ``phishing'' or ``@phish\_report''; and URLs from PhishTank, a community-based phishing verification platform.
We excluded URLs with invalid domain names (NXDOMAIN), domain parking, or legitimate site domains (matching the top 100,000 sites in the Tranco list~\cite{LePochat2019}).

\noindent\textbf{Crawling Results DB Settings.}
PhishParrot requires a sufficient number of crawling results labeled as successful crawling or failed crawling.
Therefore, we randomly selected from the aforementioned 540 crawling environments and applied them to suspicious URLs from February 1, 2025, to February 21, 2025, collecting a total of 33,432 crawling results.
We labeled these results as successful crawling (reaching phishing content) or failed crawling (content was cloaked) based on consensus among three security engineers with all engineers agreeing on each classification.
This process yielded 624 successful entries and 1,223 failed entries, which were stored in the Crawling Results DB.
Furthermore, if all Phishing Detectors judged a site to be phishing when accessed by PhishParrot, but they judged it to be non-phishing when accessed by both the Standard Analysis System and the Typical User System, we assumed a high likelihood of cloaking and also stored that crawling result in the Crawling Results DB.
This setting allows PhishParrot to capture the latest attack trends and adapt to rapidly changing phishing tactics.

\noindent\textbf{LLMs.}
PhishParrot uses o3-mini, GPT-4o, and GPT-4o mini from OpenAI.
These are the latest models with excellent inference performance and speed.
By comparing various models, we can evaluate which is optimal for adoption in PhishParrot.

\noindent\textbf{Metrics.}
We evaluate system performance using seven metrics: Accuracy, True Positive Rate, True Negative Rate, Precision, F1 score, Execution Time, and Cost, with VirusTotal as ground truth.
VirusTotal provides analysis results from approximately 90 antivirus engines but requires continuous evaluation due to detection delays for new phishing sites.
We conducted continuous monitoring for one week and adopted criteria from previous research to suppress false detections~\cite{DBLP:conf/imc/PengYS019,DBLP:conf/uss/ZhuSYQZS020}.
URLs determined as ``malicious'' by five or more engines are classified as true phishing sites; others as non-phishing sites.

Using this ground truth, we evaluate: Accuracy (Acc) as the proportion of URLs correctly classified by the system; True Positive Rate (TPR) as correct phishing classifications among true phishing sites; True Negative Rate (TNR) as correct non-phishing classifications among true non-phishing sites; Precision (Pre) as true phishing sites among system-identified phishing sites; F1 score as the harmonic mean of Precision and TPR; Execution Time as per-URL analysis time from input to output; and Cost as LLM API cost per URL (as of March 2025).

\subsection{Evaluation Results}
We conducted a real-time crawling of 32,487 suspicious URLs collected over 21 days from March 1, 2025, to March 21, 2025, using PhishParrot and two baseline systems.
We then applied three Phishing Detectors and performed continuous monitoring with VirusTotal.
According to VirusTotal's inspection results, 15,309 URLs were identified as phishing sites, while 17,178 URLs were classified as non-phishing sites.

\begin{table}[!t]
\notsotiny
\tabcolsep=1.2mm
\centering
\caption{Performance Comparison Between PhishParrot (GPT-4o mini) and Baseline Systems}
\begin{tabular}{llrrrrrr} \toprule
Systems & Phishing Detector & Acc & TPR & TNR & Pre & F1 & Execution Time \\ \midrule
PhishParrot & ChatPhishDetector & \textbf{92.7\%} & \textbf{95.0\%} & \textbf{89.6\%} & \textbf{91.8\%} & \textbf{0.933} & 26.71 seconds \\
& VisualPhishnet & 84.7\% & 86.5\% & 82.9\% & 84.5\% & 0.855 & 12.00 seconds \\
& StackModel& 86.7\% & 89.9\% & 83.5\% & 86.1\% & 0.877 & 11.37 seconds \\
\midrule
Typical & ChatPhishDetector & 64.4\% & 68.0\% & 58.1\% & 63.1\% & 0.652 & 22.15 seconds \\
User & VisualPhishnet & 52.4\% & 50.1\% & 54.8\% & 52.5\% & 0.513 & 7.44 seconds \\
System & StackModel& 54.4\% & 52.8\% & 46.0\% & 53.8\% & 0.529 & 6.81 seconds \\
\midrule
Standard & ChatPhishDetector & 58.9\% & 61.9\% & 52.4\% & 61.3\% & 0.612 & 20.31 seconds \\
Analysis & VisualPhishnet & 50.8\% & 46.4\% & 55.2\% & 50.8\% & 0.485 & 5.60 seconds \\
System & StackModel& 54.6\% & 53.9\% & 45.4\% & 53.8\% & 0.534 & 4.97 seconds \\
\bottomrule
\end{tabular}
\label{tab:performance_environment}
\end{table}

\begin{table}[!t]
\notsotiny
\tabcolsep=4.0mm
\centering
\caption{Top 5 Suggested Optimal User Profile}
\centering
\begin{tabular}{lr} \toprule
Optimal User Profile (OS, Browser, Network, Location) & \# \\ \midrule
Windows, Google Chrome, Residential, Japan & 6,128 (18.87\%) \\
Android, Google Chrome, Residential, Japan & 5,834 (17.97\%) \\
Windows, Google Chrome, Datacenter, United States & 5,492 (16.92\%) \\
macOS, Safari, Residential, Japan & 2,847 (8.77\%) \\
Android, Google Chrome, Mobile, United States & 2,432 (7.49\%) \\
  \bottomrule
\end{tabular}
\label{tab:suggested_full}
\end{table}

\begin{table}[!t]
\notsotiny
\tabcolsep=0.7mm
\centering
\caption{Top 5 Suggested User-agent and Network Type}
\begin{minipage}[b]{0.48\linewidth}
\centering
\begin{tabular}{lr} \toprule
User-agent Type & \# \\ \midrule
Windows, Google Chrome & 10,243 (31.53\%) \\
Android, Google Chrome & 9,325 (28.70\%) \\
macOS, Google Chrome & 5,543 (17.06\%) \\
macOS, Safari & 5,136 (15.81\%) \\
iOS, Safari & 853 (2.63\%) \\
  \bottomrule
\end{tabular}
\end{minipage}
\hfill
\begin{minipage}[b]{0.48\linewidth}
\centering
\begin{tabular}{lr} \toprule
 Network Type & \# \\ \midrule
 Residential, Japan & 17,487 (53.83\%) \\
 Datacenter, United States & 9,978 (30.71\%) \\
 Mobile, United States & 2,432 (7.49\%) \\
 Residential, United States & 625 (1.92\%) \\
 Mobile, Japan & 468 (1.44\%) \\
  \bottomrule
\end{tabular}
\end{minipage}
\label{tab:suggested}
\end{table}

\noindent\textbf{Summary.}
Table~\ref{tab:performance_llm} presents the performance comparison of PhishParrot across different LLMs, while Table~\ref{tab:performance_environment} compares the accuracy between PhishParrot and baseline systems.
The evaluation results reveal that GPT-4o mini demonstrates the optimal balance between detection effectiveness and processing efficiency among the compared LLMs.
Furthermore, the PhishParrot system using GPT-4o mini achieved Accuracy rates of 92.7\%, 84.7\%, and 86.7\% for ChatPhishDetector, VisualPhishnet, and StackModel, respectively.
This represents an improvement of up to 33.8\% compared to the baseline system, achieved with only an average increase of about 7.6 seconds in execution time, demonstrating PhishParrot's practicality.
These results remain valuable even if some phishing sites are inactive or employ sophisticated cloaking (e.g., advanced bot checks) requiring manual user interaction.

\noindent\textbf{LLMs.}
In evaluating PhishParrot, GPT-4o mini demonstrated the best overall performance with the highest accuracy, shortest execution time, and lowest cost. While o3-mini recorded the highest precision, it required longer processing time. GPT-4o showed the lowest accuracy and highest cost among the tested models.
The difference in detection accuracy among LLM models is within 2.1\%, which is not statistically significant.
The analysis cost per URL (\$0.002--\$0.027) is negligible for practical use, and execution time increases can be addressed through parallel processing.
Considering these results, GPT-4o mini is determined to be the optimal choice for PhishParrot.

\noindent\textbf{Baseline Systems.}
PhishParrot using GPT-4o mini significantly outperformed both baseline crawling approaches across all Phishing Detectors.
Compared to the Typical User System, PhishParrot achieved accuracy improvements of 28.3\%, 32.3\%, and 32.3\% for ChatPhishDetector, VisualPhishnet, and StackModel, respectively.
When compared to the Standard Analysis System, the difference expanded further, showing improvements of 33.8\%, 33.9\%, and 32.1\%.
This remarkable improvement demonstrates that the adaptive selection of crawling environments based on LLM analysis and similar case extraction effectively evades cloaking techniques that hinder detection.
The increase in execution time was moderate (about 5-6 seconds) compared to the dramatic improvement in detection capability.
Notably, PhishParrot showed consistent performance improvement across all detector types, demonstrating the versatility and robustness of the proposed system.

\noindent\textbf{Examples of Optimal User Profile.}
Analysis of PhishParrot's suggested user profiles revealed distinct patterns in attackers' targeting strategies.
Table~\ref{tab:suggested_full} shows the top 5 complete combinations, with Japanese users on residential networks dominating (top 2 combinations: 36.84\% combined).
Windows Google Chrome from Japanese residential networks ranked first (18.87\%), followed by Android Google Chrome from Japanese residential networks (17.97\%).
United States-based profiles also appeared prominently, with Windows Google Chrome from United States datacenters third (16.92\%).
Notably, macOS Safari from Japanese residential networks ranked fourth (8.77\%), demonstrating attackers' awareness of Apple users as valuable targets.

Individual component analysis (Table~\ref{tab:suggested}) showed Google Chrome's dominance across all platforms (77.29\% total), reflecting the attackers' focus on widely-adopted environments.
Geographic analysis revealed a strong concentration in Japanese residential networks (53.83\%) and United States datacenters (30.71\%).
The combined Japanese targeting (55.27\%) suggests attackers' particular focus on this region, potentially related to economic characteristics and cybersecurity awareness differences~\cite{gasa2024}.
These insights demonstrate PhishParrot's ability to identify sophisticated, region-specific targeting strategies.
This level of granular insight into attacker TTPs (Tactics, Techniques, and Procedures) underscores the limitations of static, rule-based analytical environments.

\section{Limitation}
PhishParrot has two main limitations.
First, the crawling environment selection process introduces additional execution overhead.
However, this overhead is offset by the improved phishing site detection rate, and the system's benefits can be maximized by selectively applying it to suspicious URLs.
Second, the system requires a certain number of successful and failed examples to reference crawling results from similar attacks.
Nevertheless, once several hundred initial examples are prepared (as in our experiment), the system begins to automatically accumulate cases of trending phishing attacks and cloaking patterns.
This adaptive mechanism is achieved through continuous operation alongside Phishing Detectors.
Therefore, after building the initial dataset, the system is designed to self-improve while adapting to the latest attack trends, maintaining PhishParrot's long-term effectiveness.

\section{Conclusion}
This research introduced PhishParrot, a crawling environment optimization system that uses LLMs to counter cloaked phishing sites.
By accumulating diverse crawling information and using LLM analysis to mimic attackers' targets, PhishParrot significantly improved phishing detection rates compared to standard analysis systems and typical user systems.
The system's ability to construct flexible environments tailored to attackers' targets demonstrates the effectiveness of combining data accumulation with LLM contextual understanding to bypass cloaking techniques.
This adaptive approach represents an advancement in phishing detection and provides a foundation for effective cybersecurity defenses against evolving threats.

{\footnotesize \bibliographystyle{unsrt}
\bibliography{bibtex}}

\begin{thebibliography}{10}

\bibitem{DBLP:conf/sp/ZhangOCSJWSKBWS21}
Penghui Zhang et~al.
\newblock Crawlphish: Large-scale analysis of client-side cloaking techniques in phishing.
\newblock In {\em Proc. IEEE SP}, 2021.

\bibitem{DBLP:conf/ccs/ZhangSKBBCO0BSA22}
Penghui Zhang et~al.
\newblock I'm spartacus, no, i'm {SPARTACUS:} proactively protecting users from phishing by intentionally triggering cloaking behavior.
\newblock In {\em Proc. ACM CCS}, 2022.

\bibitem{DBLP:conf/ccs/AbdelnabiKF20}
Sahar Abdelnabi et~al.
\newblock Visualphishnet: Zero-day phishing website detection by visual similarity.
\newblock In {\em Proc. ACM CCS}, 2020.

\bibitem{DBLP:conf/uss/LinLDNCLSZD21}
Yun Lin et~al.
\newblock Phishpedia: {A} hybrid deep learning based approach to visually identify phishing webpages.
\newblock In {\em Proc. {USENIX} Security}, 2021.

\bibitem{DBLP:conf/uss/Liu0YNDD22}
Ruofan Liu et~al.
\newblock Inferring phishing intention via webpage appearance and dynamics: {A} deep vision based approach.
\newblock In {\em Proc. {USENIX} Security}, 2022.

\bibitem{DBLP:journals/access/KoideNC24}
Takashi Koide et~al.
\newblock Chatphishdetector: Detecting phishing sites using large language models.
\newblock {\em {IEEE} Access}, 2024.

\bibitem{DBLP:conf/uss/AcharyaV21}
Bhupendra Acharya and Phani Vadrevu.
\newblock Phishprint: Evading phishing detection crawlers by prior profiling.
\newblock In {\em Proc. {USENIX} Security}, 2021.

\bibitem{DBLP:conf/uss/Liu0ZLD23}
Ruofan Liu et~al.
\newblock Knowledge expansion and counterfactual interaction for reference-based phishing detection.
\newblock In {\em Proc. {USENIX} Security}, 2023.

\bibitem{DBLP:journals/compsec/ChibaNK25}
Daiki Chiba et~al.
\newblock Domaindynamics: Advancing lifecycle-based risk assessment of domain names.
\newblock {\em Comput. Secur.}, 2025.

\bibitem{DBLP:conf/sp/InvernizziTKCPB16}
Luca Invernizzi et~al.
\newblock Cloak of visibility: Detecting when machines browse a different web.
\newblock In {\em Proc. IEEE SP}, 2016.

\bibitem{DBLP:conf/uss/Teoh0LHD24}
Xiwen Teoh et~al.
\newblock Phishdecloaker: Detecting captcha-cloaked phishing websites via hybrid vision-based interactive models.
\newblock In {\em Proc. {USENIX} Security}, 2024.

\bibitem{DBLP:conf/acsac/KoideFN023}
Takashi Koide et~al.
\newblock Phishreplicant: {A} language model-based approach to detect generated squatting domain names.
\newblock In {\em Proc. ACSAC}, 2023.

\bibitem{DBLP:conf/asiaccs/0004HK24}
Woonghee Lee et~al.
\newblock Beneath the phishing scripts: {A} script-level analysis of phishing kits and their impact on real-world phishing websites.
\newblock In {\em Proc. Asia CCS}, 2024.

\bibitem{DBLP:conf/sigir/CarbonellG98}
Jaime~G. Carbonell and Jade Goldstein.
\newblock The use of mmr, diversity-based reranking for reordering documents and producing summaries.
\newblock In {\em Proc. SIGIR}, 1998.

\bibitem{DBLP:conf/nips/LewisPPPKGKLYR020}
Patrick S.~H. Lewis et~al.
\newblock Retrieval-augmented generation for knowledge-intensive {NLP} tasks.
\newblock In {\em Proc. NeurIPS}, 2020.

\bibitem{DBLP:journals/fgcs/LiYCYL19}
Yukun Li et~al.
\newblock A stacking model using {URL} and {HTML} features for phishing webpage detection.
\newblock {\em Future Gener. Comput. Syst.}, 2019.

\bibitem{DBLP:conf/IEEEares/Nakano0KFYHYM23}
Hiroki Nakano et~al.
\newblock Canary in twitter mine: Collecting phishing reports from experts and non-experts.
\newblock In {\em Proc. ARES}, 2023.

\bibitem{LePochat2019}
Victor {Le Pochat} et~al.
\newblock Tranco: A research-oriented top sites ranking hardened against manipulation.
\newblock In {\em Proc. NDSS}, 2019.

\bibitem{DBLP:conf/imc/PengYS019}
Peng Peng et~al.
\newblock Opening the blackbox of virustotal: Analyzing online phishing scan engines.
\newblock In {\em Proc. IMC}, 2019.

\bibitem{DBLP:conf/uss/ZhuSYQZS020}
Shuofei Zhu et~al.
\newblock Measuring and modeling the label dynamics of online anti-malware engines.
\newblock In {\em Proc. {USENIX} Security}, 2020.

\bibitem{gasa2024}
Global Anti-Scam Alliance.
\newblock 2024 state of scams in japan report, 2024.
\newblock \url{https://www.gasa.org/post/2024-state-of-scams-in-japan-report}.

\end{thebibliography}

\end{document}